\documentclass[11pt]{article}
\usepackage[T1]{fontenc}
\usepackage[latin9]{inputenc}
\usepackage[a4paper]{geometry}
\usepackage{xcolor}
\usepackage{textcomp}
\usepackage{amsmath}
\usepackage{amssymb}
\usepackage{esint}

\makeatletter

\pdfoutput=0 

\usepackage{jheppub}

\usepackage{etoolbox}
    
 \patchcmd{\maketitle}{\@fpheader}{}{}{}

\usepackage{amsfonts}

\usepackage{bbm}

\title{\boldmath The (super)conformal BMS$_3$ algebra}

\author[a]{Oscar Fuentealba,}
\author[b]{Hern\'an A. Gonz\'alez,}
\author[c]{Alfredo P\'{e}rez,}
\author[d]{David Tempo}
\author[c]{and Ricardo Troncoso.}

\affiliation[a]{Universit\'e Libre de Bruxelles and International Solvay Institutes, ULB-Campus Plaine CP231, B-1050 Brussels, Belgium}
\affiliation[b]{Facultad de Artes Liberales, Universidad Adolfo Ib\'a\~{n}ez, Santiago, Chile}
\affiliation[c]{Centro de Estudios Cient\'{i}ficos (CECs), Av. Arturo Prat 514, Valdivia,
Chile}
\affiliation[d]{Departamento de Ciencias Matem\'{a}ticas y F\'{i}sicas, Universidad Cat\'olica de Temuco, Chile}

\emailAdd{ofuentea@ulb.ac.be}
\emailAdd{hernan.gonzalez@uai.cl}
\emailAdd{aperez@cecs.cl}
\emailAdd{jtempo@uct.cl}
\emailAdd{troncoso@cecs.cl}

\preprint{CECS-PHY-20/04}

\abstract{
The conformal extension of the BMS$_{3}$ algebra is constructed.
Apart from an infinite number of \textquoteleft superdilatations,\textquoteright$\,$
in order to incorporate \textquoteleft superspecial conformal transformations,\textquoteright$\,$
the commutator of the latter with supertranslations strictly requires
the presence of nonlinear terms in the remaining generators. The algebra
appears to be very rigid, in the sense that its central extensions
as well as the nonlinear terms coefficients become determined by
the central charge of the Virasoro subalgebra. The wedge algebra corresponds
to the conformal group in three spacetime dimensions $SO(3,2)$,
so that the full algebra can also be interpreted as an infinite-dimensional
nonlinear extension of the AdS$_{4}$ algebra with nontrivial central
charges. Moreover, since the Lorentz
subalgebra ($sl(2,R)$) is non-principally embedded within the conformal
(wedge) algebra, according to the conformal weight of the generators,
the conformal extension of BMS$_{3}$ can be further regarded as a
$W_{(2,2,2,1)}$ algebra. An explicit canonical realization of the
conformal extension of BMS$_{3}$ is then shown to emerge from the
asymptotic structure of conformal gravity in 3D, endowed with a new
set of boundary conditions. The supersymmetric extension is also briefly
addressed.}

\makeatother

\begin{document}
\maketitle \flushbottom

\newpage{}

\section{Introduction }

The symmetries of special relativity are embodied through the Poincar\'e
algebra. Thus, extensions thereof turn out to play a relevant role
in theoretical physics. Indeed, for relativistic systems with scale
invariance, the algebra is generically enhanced to that of the conformal
group, including special conformal transformations, see e.g. \cite{Polchinski:1987dy,Nakayama:2013is}.
Conformal field theories, formulated in terms of these enhanced symmetries,
have spanned a wealth impressive results in a wide variety of contexts
\cite{DiFrancesco:1997nk,Green:2012oqa,Cardy:1996xt,Itzykson:1988bk,Blumenhagen:2009zz}.
Besides, extensions of the Poincar\'e algebra that contain additional
fermionic generators of spin $1/2$, known as super-Poincar\'e algebras,
provide the building blocks for most of supersymmetric field theories,
enjoying a prominent and complementary source of exciting developments
\cite{Gates:1983nr,Nilles:1983ge,Haber:1984rc,West:1990tg,Martin:1997ns,Weinberg:2000cr,Binetruy:2006ad}.
Another very interesting extension of the Poincar\'e algebra, known
as the BMS algebra, emerged from the structure of asymptotically flat
spacetimes at null infinity \cite{Sachs:1962zza,Bondi:1962px}, in
which translations are enhanced to an infinite-dimensional ideal of
``supertranslations''. The BMS algebra can be further extended to
admit ``superrotations'' \cite{Barnich:2009se,Barnich:2010eb,Barnich:2011ct,Campiglia:2014yka,Campiglia:2015yka}
and it has recently attracted a great deal of attention due to its
fascinating connections with soft theorems \cite{Low:1954kd,Weinberg:1965nx,He:2014laa},
the memory effect \cite{Strominger:2014pwa}, and the information
paradox \cite{Hawking:2016msc,Hawking:2016sgy}. More recently, the
robustness of the BMS algebra shows itself through its canonical realization
either at null \cite{Bunster:2018yjr} or spatial infinity \cite{Henneaux:2018cst,Henneaux:2019yax,Fuentealba:2020ghw},
and also near generic horizons \cite{Grumiller:2019fmp}.

It is then natural to wonder about the possible compatibility of these
three time-honored, but wildly different extensions of the Poincar\'e
algebra.

Conformal and supersymmetric extensions of the Poincar\'e algebra turn
out to be perfectly compatible through the well-known superconformal
algebra \cite{Haag:1974qh,Gates:1983nr}. Nevertheless, the supersymmetric
extension of the BMS algebra remains intriguing. Indeed, among the
infinite number of supertranslations, only the subset of standard
translations possesses a fermionic ``square root'' being spanned
by four fermionic generators, at null \cite{Awada:1985by} or spatial
infinity \cite{Henneaux:2020ekh}. Inequivalent extensions with an
infinite number of fermionic generators have been proposed in \cite{Awada:1985by,Avery:2015iix,Fotopoulos:2020bqj},
and it is still unclear whether they could be canonically realized
even at the linearized level \cite{Fuentealba:2020aax}.

On the other hand, a conformal extension of BMS has been recently
constructed in \cite{Haco:2017ekf}, which successfully accommodates
\textquoteleft superdilatations\textquoteright . However, its structure
is very different from that of the conformal group since standard
special conformal transformations are not included. Thus, the BMS
algebra seems to resist compatibility with the full conformal extension.

In the case of three-dimensional spacetimes, the conformal, supersymmetric
and BMS extensions of the Poincar\'e algebra are also well-known. The
compatibility of conformal and minimal supersymmetric extensions is
also firmly established by the superconformal algebra $osp(1|4)$
\cite{Nahm:1977tg}. Interestingly, in contradistinction to the four-dimensional
case, the BMS$_{3}$ algebra \cite{Ashtekar:1996cd,Barnich:2006av}
is known to admit a fully-fledged supersymmetric extension, in the
sense that supertranslations possess suitable fermionic square roots,
spanned by an infinite number of fermionic canonical generators \cite{Barnich:2014cwa}.
However, as in four dimensions, a full conformal extension of BMS$_{3}$
has not been hitherto reported. In fact, as it has been recently pointed
out from entirely different approaches in \cite{Adami:2020ugu}, \cite{Donnay:2020fof},
\cite{Batlle:2020hia} the BMS$_{3}$ algebra can be suitably enlarged
by superdilatations, but nonetheless, some difficulties in the closure
of the algebra seem to preclude the inclusion of special conformal
transformations. 

\section{The conformal BMS$_{3}$ algebra\label{ConfBMS3-alg}}

Here we show that the conformal extension of the BMS$_{3}$ algebra
that incorporates \textquoteleft superspecial conformal transformations\textquoteright{}
is a nonlinear algebra. In particular, the commutator of supertranslations
with special conformal transformations strictly requires the presence
of nonlinear terms in the remaining generators, which acquire support
provided that the \textquoteleft BMS$_{3}$-Weyl\textquoteright{}
subalgebra is endowed with nontrivial central extensions. This can
be seen as follows.

It is simple to verify that the BMS$_{3}$ algebra, spanned by superrotations
${\cal J}_{m}$ and supertranslations ${\cal P}_{m}$, once enlarged
by superdilatations ${\cal D}_{m}$, admits only two nontrivial central
extensions. The centrally-extended BMS$_{3}$-Weyl algebra then reads
\begin{align}
i\left\{ \mathcal{J}_{m},\mathcal{J}_{n}\right\}  & =\left(m-n\right)\mathcal{J}_{m+n}+c\left(m^{2}-1\right)m\delta_{m+n,0}\,,\nonumber \\
i\left\{ \mathcal{J}_{m},\mathcal{P}_{n}\right\}  & =\left(m-n\right)\mathcal{P}_{m+n}\,,\nonumber \\
i\left\{ \mathcal{J}_{m},\mathcal{D}_{n}\right\}  & =-n\mathcal{D}_{m+n}\,,\label{eq:Weyl-BMS3-algebra}\\
i\left\{ \mathcal{P}_{m},\mathcal{D}_{n}\right\}  & =-i\mathcal{P}_{m+n}\,,\nonumber \\
i\left\{ \mathcal{D}_{m},\mathcal{D}_{n}\right\}  & =\tilde{c}m\delta_{m+n,0}\,,\nonumber 
\end{align}
where $m$, $n\in\mathbb{Z}$. Vanishing commutators are omitted hereafter.
Note that in presence of superdilatations, the Jacobi identity excludes
the possibility of a nontrivial central charge in the commutator of
${\cal J}_{m}$ and ${\cal P}_{n}$.

The generators of superspecial conformal transformations ${\cal K}_{m}$
can then be incorporated provided that the superdilatations ``level''
$\tilde{c}$ coincides with the central charge of the Virasoro subalgebra
($\tilde{c}=c$), so that the remaining commutators of the full conformal
BMS$_{3}$ algebra are given by 
\begin{align}
i\left\{ {\cal J}_{m},{\cal K}_{n}\right\}  & =\left(m-n\right)\mathcal{K}_{m+n}\,,\nonumber \\
i\left\{ {\cal K}_{m},{\cal D}_{n}\right\}  & =i\mathcal{K}_{m+n}\,,\label{BMS3-Conf-Algebra-mn}\\
i\left\{ {\cal P}_{m},{\cal K}_{n}\right\}  & =-2\left(m-n\right)\mathcal{J}_{m+n}-2i\left(m^{2}-mn+n^{2}-1\right)\mathcal{D}_{m+n}\nonumber \\
 & +\left(m-n\right)\Lambda_{m+n}^{\left(2\right)}+\Lambda_{m+n}^{\left(3\right)}-2c\left(m^{2}-1\right)m\delta_{m+n,0}\,,\nonumber 
\end{align}
where $\Lambda_{m}^{\left(s\right)}$ stands for nonlinear terms defined
through
\begin{align}
\Lambda_{m}^{\left(2\right)} & =\frac{4}{c}\sum_{n}{\cal D}_{m-n}{\cal D}_{n}\;,\\
\Lambda_{m}^{\left(3\right)} & =-\frac{4i}{c}\sum_{n}{\cal J}_{m-n}{\cal D}_{n}+\frac{4i}{c^{2}}\sum_{n,l}{\cal D}_{m-n-l}{\cal D}_{n}{\cal D}_{l}\;,\label{Lambda3}
\end{align}
with (anomalous) conformal weight $s$. Indeed, the conformal weight
of ${\cal J}_{m}$, ${\cal P}_{m}$ and ${\cal K}_{m}$ is given by
$s=2$, while ${\cal D}_{m}$ has conformal weight $s=1$. 

It is worth highlighting that the central extensions as well as the
coefficients at front of the nonlinear terms of the conformal BMS$_{3}$
algebra turn out to be entirely determined by the central charge $c$
of the Virasoro subalgebra, and in this sense, the algebra is very
rigid. 

The wedge algebra reduces to that of the conformal group $SO(3,2)$.
It is recovered by restricting the integers that label the generators
according to their conformal weight $s$ as $\left|m\right|<s$, dropping
nonlinear terms (see Eqs. \eqref{eq:so(3,2)Wedge} and \eqref{eq:WedgeModesQ}).

Noteworthy, the conformal BMS$_{3}$ algebra can then also be interpreted
as an infinite-dimensional nonlinear extension of the AdS$_{4}$ algebra
with nontrivial central charges. In this way, the classical theorem
of algebraic cohomology that precludes nontrivial central extensions
for semisimple algebras (see e.g. \cite{Fuks}), clearly does not
apply in this case due to the nonlinearity of the extended algebra.

Furthermore, as the Lorentz subalgebra ($sl(2,\mathbb{R})$), spanned
by ${\cal J}_{m}$ with $m=-1,0,1$, is non-principally embedded within
the wedge algebra ($so(3,2)$), taking into account the conformal
weight of the generators, the conformal extension of BMS$_{3}$ can
also be regarded as a $W_{(2,2,2,1)}$ algebra (see e.g. \cite{Bouwknegt:1995ag,Frappat:1992bs})\footnote{An isomorphism between the pure BMS$_{3}$ algebra and the so-called
W$(2,2)$ algebra has also been pointed out in \cite{Rasmussen:2017eus},
and further elaborated in \cite{Aizawa-Kimura,Adamovic-Radobolja1,Zhang-Dong,Jiang-Zhang,Adamovic-Radobolja2,Araujo:2018dem}.}.

It is also worth pointing out that, as it occurs for classical $W$-algebras,
the conformal BMS$_{3}$ algebra is well defined provided that the
Virasoro central charge does not vanish; since otherwise, the coefficients
that give support to the nonlinear terms would blow up. Nevertheless,
this is not necessarily the case for the quantum algebra because these
coefficients as well as the central extensions generically acquire
corrections. 

An explicit canonical realization of the conformal BMS$_{3}$ algebra
is performed in the next section, while the superconformal extension
of BMS$_{3}$ is briefly addressed in section \textcolor{brown}{\ref{eq:SuperconformalBMS3}}.

\section{Explicit realization: asymptotic structure of conformal gravity in
3D\label{Canonical}}

The aforementioned link between the conformal BMS$_{3}$ and $W_{(2,2,2,1)}$
algebras, naturally suggests an explicit realization in terms of a
WZW model for $SO(3,2)$ \cite{Frappat:1992bs}, so that the conformal
BMS$_{3}$ algebra could be obtained from the Kac-Moody extension
of $so(3,2)$ by virtue of a Sugawara-like construction\footnote{Indeed, this sort of Sugawara-like construction can be successfully
implemented in order to obtain the (super) BMS$_{3}$ algebra from
the affine extension of (super) Poincar\'e in 3D \cite{Barnich:2013yka,Barnich:2015sca}
(see also \cite{Banerjee:2019lrv}). }. The Kac-Moody currents could also be endowed with suitable constraints
so that the conformal BMS$_{3}$ algebra emerges from the Dirac brackets.
The latter option can be holographically realized along the lines
of \cite{Coussaert:1995zp}, so that the constraints would be automatically
implemented through an appropriate choice of boundary conditions for
a Chern-Simons theory of $SO(3,2)$.

As shown in \cite{Horne:1988jf}, a Chern-Simons theory for $SO(3,2)$,
described by
\begin{equation}
I_{CS}\left[A\right]=\frac{k}{4\pi}\int\left\langle AdA+\frac{2}{3}A^{3}\right\rangle \thinspace,\label{eq:ICS}
\end{equation}
turns out to be related to conformal gravity in 3D \cite{Deser:1981wh,Deser:1982vy},
which admits an interesting class of black hole solutions \cite{Oliva:2009hz}
(see also \cite{Oliva:2009ip}). Some choices of asymptotic conditions
for conformal gravity in 3D have already been explored in \cite{Afshar:2011qw,Bertin:2012qw,Afshar:2013bla},
being such that the asymptotic symmetry algebra is given by the direct
sum of a $U\left(1\right)$ current with either BMS$_{3}$ or two
copies of the Virasoro algebra. Nevertheless, these choices do not
accommodate the black holes in \cite{Oliva:2009hz}. Thus, in what
follows we propose a new set of boundary conditions that allows to
include them, and also provides a canonical realization of the conformal
BMS$_{3}$ algebra that emerges from the asymptotic symmetries. 

Following \cite{Coussaert:1995zp}, the radial dependence of the asymptotic
form of the gauge field can be completely gauged away by virtue of
a gauge choice of the form $A=g^{-1}ag+g^{-1}dg$, with $g=g\left(r\right)$,
so that the components of the auxiliary connection $a=a_{t}dt+a_{\varphi}d\varphi$
depend only on time and the angular coordinate.

It is useful to express the generators of $SO\left(3,2\right)$ in
a basis that matches that of the wedge algebra described in Section
\ref{ConfBMS3-alg}, being precisely defined in \eqref{eq:so(3,2)Wedge}.
Thus, the asymptotic behavior that we propose for $a_{\varphi}$ can
be readily written in terms of deviations with respect to a reference
configuration that go along highest weight generators, i.e., 
\begin{align}
a & _{\varphi}=J_{1}-\frac{\pi}{k}\left(\left(\mathcal{J}-\frac{\pi}{k}\mathcal{D}^{2}\right)J_{-1}+\frac{1}{2}\mathcal{P}P_{-1}+\frac{1}{2}\mathcal{K}K_{-1}-2\mathcal{D}D\right)\;,\label{eq:a_phi}
\end{align}
where the dynamical fields ${\cal J}$, ${\cal P}$, ${\cal K}$,
${\cal D}$ depend on $t$, $\varphi$. This fall-off is maintained
under gauge transformations $\delta a=d\Omega+\left[a,\Omega\right]$,
where $\Omega=\Omega\left[\epsilon_{{\cal J}},\epsilon_{{\cal P}},\epsilon_{{\cal K}},\epsilon_{{\cal D}}\right]$
depends on four arbitrary functions of $t$, $\varphi$ ($\epsilon_{{\cal X}}=\epsilon_{{\cal X}}\left(t,\varphi\right)$).
The explicit form of $\Omega$ as well as the transformation law of
the dynamical fields are given in Eqs. \eqref{eq:Omega} and \eqref{eq:Tranflaws},
respectively. According to \cite{Henneaux:2013dra,Bunster:2014mua},
the asymptotic symmetries are preserved by the evolution in time by
choosing the asymptotic form of $a_{t}$ to be generically given by
\begin{equation}
a_{t}=\Omega\left[\mu_{{\cal J}},\mu_{{\cal P}},\mu_{{\cal K}},\mu_{{\cal D}}\right]\thinspace,\label{eq: a_t}
\end{equation}
where the ``chemical potentials'' $\mu_{{\cal X}}=\mu_{{\cal X}}\left(t,\varphi\right)$
are assumed to be fixed at the boundary. The fall-off of $a_{t}$
is then maintained by the asymptotic symmetries provided that the
field equations hold in the asymptotic region, and the parameters
$\epsilon_{{\cal X}}$ fulfill suitable differential equations of
first order in time (see \eqref{eq:Epsilon-Punto}).

The asymptotic symmetry generators can then be obtained from different
approaches \cite{Regge:1974zd,Barnich:2001jy}, which read
\begin{equation}
\mathcal{Q}\left[\epsilon_{{\cal J}},\epsilon_{{\cal P}},\epsilon_{{\cal K}},\epsilon_{{\cal D}}\right]=-\int\left(\epsilon_{{\cal J}}\mathcal{J}+\epsilon_{{\cal P}}\mathcal{P}+\epsilon_{{\cal K}}\mathcal{K}+\epsilon_{{\cal D}}\mathcal{D}\right)d\varphi\,.\label{eq:Q}
\end{equation}
The algebra of the conserved charges \eqref{eq:Q} can then be obtained
from their Dirac brackets, or more directly from the transformation
law of the fields in Eq. \eqref{eq:Tranflaws} by virtue of $\left\{ {\cal Q}\left[\eta_{1}\right],{\cal Q}\left[\eta_{2}\right]\right\} =-\delta_{\eta_{1}}{\cal Q}\left[\eta_{2}\right]$.
It is explicitly given by 
\begin{align}
\left\{ {\cal J}\left(\phi\right),{\cal J}\left(\varphi\right)\right\}  & =-2\mathcal{J}\left(\phi\right)\delta^{\prime}\left(\phi-\varphi\right)-\delta\left(\phi-\varphi\right)\mathcal{J}^{\prime}\left(\phi\right)+\frac{k}{2\pi}\delta^{\prime\prime\prime}\left(\phi-\varphi\right)\;,\nonumber \\
\left\{ {\cal J}\left(\phi\right),{\cal P}\left(\varphi\right)\right\}  & =-2\mathcal{P}\left(\phi\right)\delta^{\prime}\left(\phi-\varphi\right)-\delta\left(\phi-\varphi\right)\mathcal{P}^{\prime}\left(\phi\right)\;,\nonumber \\
\left\{ {\cal J}\left(\phi\right),{\cal K}\left(\varphi\right)\right\}  & =-2\mathcal{K}\left(\phi\right)\delta^{\prime}\left(\phi-\varphi\right)-\delta\left(\phi-\varphi\right)\mathcal{K}^{\prime}\left(\phi\right)\;,\nonumber \\
\left\{ {\cal J}\left(\phi\right),{\cal D}\left(\varphi\right)\right\}  & =-\mathcal{D}\left(\phi\right)\delta^{\prime}\left(\phi-\varphi\right)\;,\nonumber \\
\left\{ {\cal P}\left(\phi\right),{\cal D}\left(\varphi\right)\right\}  & =-{\cal P}\left(\phi\right)\delta\left(\phi-\varphi\right)\;,\label{eq:PoissonBracBMS3Conf}\\
\left\{ {\cal K}\left(\phi\right),{\cal D}\left(\varphi\right)\right\}  & ={\cal K}\left(\phi\right)\delta\left(\phi-\varphi\right)\;,\nonumber \\
\left\{ {\cal D}\left(\phi\right),{\cal D}\left(\varphi\right)\right\}  & =-\frac{k}{2\pi}\delta^{\prime}\left(\phi-\varphi\right)\;,\nonumber \\
\left\{ {\cal P}\left(\phi\right),{\cal K}\left(\varphi\right)\right\}  & =4\left(\mathcal{J}\left(\phi\right)-\frac{4\pi}{k}\mathcal{D}^{2}\left(\phi\right)\right)\delta^{\prime}\left(\phi-\varphi\right)+2\left(\mathcal{J}\left(\phi\right)-\frac{4\pi}{k}\mathcal{D}^{2}\left(\phi\right)\right)^{\prime}\delta\left(\phi-\varphi\right)\nonumber \\
 & +2\left[\mathcal{D}^{\prime\prime}\left(\phi\right)-\frac{4\pi}{k}\left(\mathcal{J}\left(\phi\right)-\frac{2\pi}{k}\mathcal{D}^{2}\left(\phi\right)\right)\mathcal{D}\left(\phi\right)\right]\delta\left(\phi-\varphi\right)\nonumber \\
 & +6\left(\mathcal{D}\left(\phi\right)\delta^{\prime}\left(\phi-\varphi\right)\right)^{\prime}-\frac{k}{\pi}\delta^{\prime\prime\prime}\left(\phi-\varphi\right)\;,\nonumber 
\end{align}
so that once expanded in Fourier modes, $X=\frac{1}{2\pi}\sum_{m}X_{m}e^{im\varphi}$,
it reduces to that in Eqs. \eqref{eq:Weyl-BMS3-algebra} and \eqref{BMS3-Conf-Algebra-mn},
with $\tilde{c}=c=k$, provided that the zero mode of ${\cal J}_{n}$
is shifted as ${\cal J}_{0}\rightarrow{\cal J}_{0}-\frac{k}{4\pi}$.

It is worth highlighting that the central extensions of the conformal
BMS$_{3}$ algebra, in this context are determined by the Chern-Simons
level $k$. This goes by hand with the fact that the conformal group
$SO(3,2)$ is semisimple, and hence, it admits a unique invariant
bilinear form being given by the Cartan-Killing metric (up to a normalization
that can fixed as in \eqref{eq:bilinearForm}).

A remarkable fact of the asymptotic behavior described above is that,
since it accommodates the black holes in \cite{Oliva:2009hz}, it
includes asymptotically (A)dS or flat three-dimensional spacetimes.
Indeed, the precise value of the ``cosmological constant'' can be
seen to be fixed by a suitable quotient of the chemical potentials.
The structure of the generic form of the black holes that fit within
our asymptotic conditions turns out to be very rich and intricate,
and it can be carefully analyzed in terms of the conserved charges
that span the conformal BMS$_{3}$ algebra. This is left for a forthcoming
work.

\section{The superconformal BMS$_{3}$ algebra\label{superConfBMS3}}

The conformal, supersymmetric and BMS extensions of the Poincar\'e algebra
in 3D can also be seen fully compatible. Indeed, the fermionic generators
of the superconformal algebra $osp\left(1|4\right)$, associated to
the square roots of translations ($Q$) and special conformal transformations
($S$), admit infinite-dimensional extensions that we denote by $\psi_{m}^{\left[+\right]}$
and $\psi_{m}^{\left[-\right]}$, corresponding to the square roots
of supertranslations and superspecial conformal transformations, respectively.

The superconformal BMS$_{3}$ algebra is then spanned by the set (${\cal J}_{m}$,
${\cal P}_{m}$, ${\cal D}_{m}$, ${\cal K}_{m}$, $\psi_{m}^{\left[+\right]}$,
$\psi_{m}^{\left[-\right]}$), so that the commutators of the BMS$_{3}$-Weyl
subalgebra $({\cal J}_{m},{\cal P}_{m},{\cal D}_{m})$ are given by
\eqref{eq:Weyl-BMS3-algebra} with $\tilde{c}=c$; while the commutators
of the generators of superspecial conformal transformations (${\cal K}_{m}$)
with the remaining bosonic generators read as in \eqref{BMS3-Conf-Algebra-mn},
where the nonlinear term of conformal weight 3 in \eqref{Lambda3}
acquires a quadratic shift in the fermionic generators, according
to

\begin{equation}
\Lambda_{m}^{\left(3\right)}\rightarrow\Lambda_{m}^{\left(3\right)}+\frac{2i}{c}\sum_{n}\psi_{m-n}^{\left[-\right]}\psi_{n}^{\left[+\right]}\;.
\end{equation}
The (anti-)commutators that involve fermionic generators read as
\begin{align}
i\left\{ {\cal J}_{m},\psi_{n}^{\left[\pm\right]}\right\}  & =\left(\frac{m}{2}-n\right)\psi_{n}^{\left[\pm\right]}\,,\nonumber \\
i\left\{ {\cal D}_{m},\psi_{n}^{\left[\pm\right]}\right\}  & =\pm\frac{i}{4}\psi_{n}^{\left[\pm\right]}\,,\nonumber \\
i\left\{ {\cal P}_{m},\psi_{n}^{\left[-\right]}\right\}  & =2\left(\frac{m}{2}-n\right)\psi_{n}^{\left[+\right]}+\Lambda_{m+n}^{\left[+\right]\left(5/2\right)}\,,\nonumber \\
i\left\{ {\cal K}_{m},\psi_{n}^{\left[+\right]}\right\}  & =-2\left(\frac{m}{2}-n\right)\psi_{n}^{\left[-\right]}-\Lambda_{m+n}^{\left[-\right]\left(5/2\right)}\,,\label{eq:SuperconformalBMS3}\\
i\left\{ \psi_{m}^{\left[+\right]},\psi_{n}^{\left[+\right]}\right\}  & ={\cal P}_{m+n}\;,\nonumber \\
i\left\{ \psi_{m}^{\left[-\right]},\psi_{n}^{\left[-\right]}\right\}  & =-{\cal K}_{m+n}\;,\nonumber \\
i\left\{ \psi_{m}^{\left[+\right]},\psi_{n}^{\left[-\right]}\right\}  & ={\cal J}_{m+n}-i\left(m-n\right){\cal D}_{m+n}-\frac{1}{4}\Lambda_{m+n}^{\left(2\right)}+2c\left(m^{2}-\frac{1}{4}\right)\delta_{m+n,0}\;,\nonumber 
\end{align}
where $\Lambda_{m}^{\left[\pm\right]\left(5/2\right)}=\frac{2i}{c}\sum_{n}{\cal D}_{m-n}\psi_{n}^{\left[\pm\right]}\;,$
and the brackets between fermionic generators are symmetric. The fermionic
generators are labelled by integers or half-integers for fermionic
parameters with periodic or antiperiodic boundary conditions, respectively. 

Note that the conformal weight of the fermionic generators $\psi_{m}^{\left[\pm\right]}$
is given by $s=3/2$. For antiperiodic boundary conditions, the wedge
algebra reduces to $osp\left(1|4\right)$, being recovered once nonlinear
terms are dropped and the labels of the generators are restricted
according to $\left|m\right|<s$, where $s$ is their conformal weight.
Therefore, the conformal weight of the generators of the superconformal
BMS$_{3}$ algebra, naturally suggests that it could be regarded as
a $W_{(2,2,2,\frac{3}{2},\frac{3}{2},1)}$ algebra.

As in the bosonic case, the superalgebra also appears to be very rigid,
in the sense that the coefficients that characterize the nonlinear
terms and all of the central extensions become completely determined
by the central charge of the Virasoro subalgebra.

It is also worth pointing out that the superconformal extension of
the BMS$_{3}$ algebra can be interpreted as an infinite-dimensional
centrally-extended nonlinear extension of the super AdS$_{4}$ algebra
($osp\left(1|4\right)$), suggesting the possibility of a different
version of the AdS$_{4}$/CFT$_{3}$ correspondence \cite{Aharony:2008ug},
presumably topological and with enhanced symmetries.

A canonical realization of the superconformal BMS$_{3}$ algebra can
also be seen to arise from the asymptotic structure of conformal supergravity
in 3D \cite{vanNieuwenhuizen:1985cx,Rocek:1985bk}, by virtue of a
suitable supersymmetric extension of the new boundary conditions presented
in section \ref{Canonical} (work in progress).

As a final remark, it might be interesting to explore whether the
super BMS$_{3}$ algebras with ${\cal N}>1$ in \cite{Banerjee:2016nio,Lodato:2016alv,Banerjee:2017gzj,Basu:2017aqn,Fuentealba:2017fck,Poojary:2017xgn},
as well as the bosonic and fermionic higher spin extensions of BMS$_{3}$
in \cite{Afshar:2013vka,Gonzalez:2013oaa,Gary:2014ppa,Matulich:2014hea}
and \cite{Fuentealba:2015jma,Fuentealba:2015wza}, respectively, could
also be compatible with the conformal extension developed here. The
compatibility with other possible extensions of BMS$_{3}$ as in e.g.,
\cite{Caroca:2017onr,Caroca:2019dds} also deserves attention.

\medskip{}

\acknowledgments 

We thank Marcela C\'ardenas, Ricardo Caroca, Joaquim Gomis, Marc Henneaux,
Javier Matulich, F\'abio Novaes, Miguel Pino, and Pablo Rodr\'iguez for
useful discussions. O.F. holds a \textquotedblleft Marina Solvay\textquotedblright{}
fellowship. This work was partially supported by the ERC Advanced
Grant \textquotedblleft High-Spin-Grav\textquotedblright , by FNRS-Belgium
(conventions FRFC PDRT.1025.14 and IISN 4.4503.15). This research
has been partially supported by FONDECYT grants N\textdegree\, 1171162, 1181031,
1181496, 11190427. The Centro de Estudios Cient\'ificos (CECs) is funded
by the Chilean Government through the Centers of Excellence Base Financing
Program of Conicyt.

\appendix

\section{Remarks on $so\left(3,2\right)$ and the conformal BMS$_{3}$ algebra\label{Appendix}}

The $so\left(3,2\right)$ algebra, spanned by generators $J_{AB}$,
which reads
\begin{equation}
\left[J_{AB},J_{CD}\right]=\eta_{AC}J_{BD}-\eta_{BC}J_{AD}+\eta_{AD}J_{CB}-\eta_{BD}J_{CA},\label{eq:JAB-algbra}
\end{equation}
 is well-known to be isomorphic to the conformal algebra in 3D. Indeed,
choosing $\eta_{AB}=diag\left(-1,1,1,1,-1\right)$ and splitting the
index $A$ according to $A=\{a,3,4\}$, the following change of basis
\begin{equation}
J_{a}=\frac{1}{2}\epsilon_{abc}J^{bc},\quad P_{a}=J_{a3}-J_{a4},\quad K_{a}=J_{a3}+J_{a4},\quad D=J_{34},
\end{equation}
makes the algebra in \eqref{eq:JAB-algbra} to read as

\begin{eqnarray}
 &  & \left[J_{a},J_{b}\right]=\epsilon_{abc}J^{c},\quad\left[P_{a},J_{b}\right]=\epsilon_{abc}P^{c},\quad\left[K_{a},J_{b}\right]=\epsilon_{abc}K^{c},\\
 &  & \left[P_{a},D\right]=P_{a},\quad\left[K_{a},D\right]=-K_{a},\quad\left[P_{a},K_{b}\right]=-2\epsilon_{abc}J^{c}+2\eta_{ab}D.
\end{eqnarray}
Therefore, if the Cartan-Killing metric is normalized according to
\begin{equation}
\left\langle J^{AB},J_{CD}\right\rangle =-\delta_{CD}^{AB}\;,
\end{equation}
in the ``conformal basis'' the nonvanishing components of the invariant
bilinear metric are given by

\begin{equation}
\left\langle J_{a},J_{b}\right\rangle =\eta_{ab}\;\;;\;\;\left\langle P_{a},K_{b}\right\rangle =-2\eta_{ab}\;\;;\;\;\left\langle D,D\right\rangle =1\;.\label{eq:bilinearForm}
\end{equation}
Besides, $so\left(3,2\right)$ also corresponds to the wedge algebra
of the conformal extension of BMS$_{3}$. In order to see that explicitly,
it is useful to choose the Minkowski metric $\eta_{ab}$ in light-cone
coordinates, so that its nonvanishing components read $\eta_{01}=\eta_{10}=\eta_{22}=1$,
and an orientation for which the Levi-Civita symbol fulfills $\epsilon_{012}=1$.
The suitable change of basis can then be defined as

\begin{align}
J_{0} & \rightarrow-\frac{1}{2}J_{-1}\;;\;J_{2}\rightarrow J_{0}\;,\\
P_{0} & \rightarrow-\frac{1}{2}P_{-1}\;;\;P_{2}\rightarrow P_{0}\;,\\
K_{0} & \rightarrow-\frac{1}{2}K_{-1}\;;\;K_{2}\rightarrow K_{0}\;,
\end{align}
and hence, the $so\left(3,2\right)$ algebra in the new basis spanned
by ($J_{m}$,$P_{n}$,$K_{m}$,$D$), with $m,n=-1,0,1$, reduces
to

\begin{align}
\left[J_{m},J_{n}\right] & =\left(m-n\right)J_{m+n}\,,\nonumber \\
\left[J_{m},P_{n}\right] & =\left(m-n\right)P_{m+n}\,,\nonumber \\
\left[J_{m},K_{n}\right] & =\left(m-n\right)K_{m+n}\,,\nonumber \\
\left[P_{m},D\right] & =P_{m}\,,\label{eq:so(3,2)Wedge}\\
\left[K_{m},D\right] & =-K_{m}\,,\nonumber \\
\left[P_{m},K_{n}\right] & =-2\left(m-n\right)J_{m+n}-2\left(m^{2}-mn+n^{2}-1\right)D\;,\nonumber 
\end{align}
which agrees with the wedge algebra of the conformal BMS$_{3}$ algebra
provided that $i\left\{ \;,\;\right\} \rightarrow\left[\;,\;\right]$,
and
\begin{align}
\mathcal{J}_{m} & \rightarrow J_{m},\;\mathcal{P}_{m}\rightarrow P_{m}\;,\mathcal{K}_{m}\rightarrow K_{m}\;,i{\cal D}_{0}\rightarrow D\;.\label{eq:WedgeModesQ}
\end{align}
In the basis \eqref{eq:so(3,2)Wedge}, the $so\left(3,2\right)$-valued
parameter $\Omega$ that preserves the asymptotic form of the gauge
field $a_{\varphi}$ in \eqref{eq:a_phi} is given by
\begin{equation}
\Omega\left[\epsilon_{{\cal J}},\epsilon_{{\cal P}},\epsilon_{{\cal K}},\epsilon_{{\cal D}}\right]=\epsilon_{{\cal J}}J_{1}-\epsilon_{{\cal K}}P_{1}-\epsilon_{{\cal P}}K_{1}+\left(\epsilon_{{\cal D}}+\frac{2\pi}{k}{\cal D}\epsilon_{{\cal J}}\right)D+\eta\left[\epsilon_{{\cal J}},\epsilon_{{\cal P}},\epsilon_{{\cal K}},\epsilon_{{\cal D}}\right]\thinspace,\label{eq:Omega}
\end{equation}
with
\begin{align}
\eta\left[\epsilon_{{\cal J}},\epsilon_{{\cal P}},\epsilon_{{\cal K}},\epsilon_{{\cal D}}\right] & =-\epsilon_{{\cal J}}^{\prime}J_{0}+\left(\epsilon_{{\cal K}}^{\prime}-\frac{2\pi}{k}{\cal D}\epsilon_{{\cal K}}\right)P_{0}+\left(\epsilon_{{\cal P}}^{\prime}-\frac{2\pi}{k}{\cal D}\epsilon_{{\cal P}}\right)K_{0}\nonumber \\
 & -\frac{\pi}{k}\left(\left({\cal J}-\frac{\pi}{k}{\cal D}^{2}\right)\epsilon_{{\cal J}}+{\cal K}\epsilon_{{\cal K}}+{\cal P}\epsilon_{{\cal P}}-\frac{k}{2\pi}\epsilon_{{\cal J}}^{\prime\prime}\right)J_{-1}\nonumber \\
 & +\frac{\pi}{k}\left(\left({\cal J}-\frac{3\pi}{k}{\cal D}^{2}+{\cal D}^{\prime}\right)\epsilon_{{\cal K}}+2{\cal D}\epsilon_{{\cal K}}^{\prime}-\frac{1}{2}{\cal P}\epsilon_{{\cal J}}-\frac{k}{2\pi}\epsilon_{{\cal K}}^{\prime\prime}\right)P_{-1}\label{eq:eta}\\
 & +\frac{\pi}{k}\left(\left({\cal J}-\frac{3\pi}{k}{\cal D}^{2}-{\cal D}^{\prime}\right)\epsilon_{{\cal P}}-2{\cal D}\epsilon_{{\cal P}}^{\prime}-\frac{1}{2}{\cal K}\epsilon_{{\cal J}}-\frac{k}{2\pi}\epsilon_{{\cal P}}^{\prime\prime}\right)K_{-1}\;,\nonumber 
\end{align}
so that the transformation law of the dynamical fields reads
\begin{align}
\delta\mathcal{J} & =2\mathcal{J}\epsilon_{{\cal J}}^{\prime}+\mathcal{J}^{\prime}\epsilon_{{\cal J}}-\frac{k}{2\pi}\epsilon_{{\cal J}}^{\prime\prime\prime}+2\mathcal{P}\epsilon_{{\cal P}}^{\prime}+\mathcal{P}^{\prime}\epsilon_{{\cal P}}+2\mathcal{K}\epsilon_{{\cal K}}^{\prime}+\mathcal{K}^{\prime}\epsilon_{{\cal K}}+\mathcal{D}\epsilon_{{\cal D}}^{\prime}\thinspace,\nonumber \\
\delta\mathcal{P} & =2\mathcal{P}\epsilon_{{\cal J}}^{\prime}+\mathcal{P}^{\prime}\epsilon_{{\cal J}}-4\left(\mathcal{J}-\frac{4\pi}{k}{\cal D}^{2}\right)\epsilon_{{\cal K}}^{\prime}-2\left(\mathcal{J}-\frac{4\pi}{k}{\cal D}^{2}\right)^{\prime}\epsilon_{{\cal K}}+\frac{k}{\pi}\epsilon_{{\cal K}}^{\prime\prime\prime}\nonumber \\
 & -2\left(\mathcal{D}^{\prime\prime}-\frac{4\pi}{k}\left(\mathcal{J}-\frac{2\pi}{k}{\cal D}^{2}\right){\cal D}\right)\epsilon_{{\cal K}}-6\left(\mathcal{D}\epsilon_{{\cal K}}^{\prime}\right)^{\prime}+\mathcal{P}\epsilon_{{\cal D}}\thinspace,\nonumber \\
\delta\mathcal{K} & =2\mathcal{K}\epsilon_{{\cal J}}^{\prime}+\mathcal{K}^{\prime}\epsilon_{{\cal J}}-4\left(\mathcal{J}-\frac{4\pi}{k}{\cal D}^{2}\right)\epsilon_{{\cal P}}^{\prime}-2\left(\mathcal{J}-\frac{4\pi}{k}{\cal D}^{2}\right)^{\prime}\epsilon_{{\cal P}}+\frac{k}{\pi}\epsilon_{{\cal P}}^{\prime\prime\prime}\nonumber \\
 & +2\left(\mathcal{D}^{\prime\prime}-\frac{4\pi}{k}\left(\mathcal{J}-\frac{2\pi}{k}{\cal D}^{2}\right){\cal D}\right)\epsilon_{{\cal P}}+6\left(\mathcal{D}\epsilon_{{\cal P}}^{\prime}\right)^{\prime}-\mathcal{K}\epsilon_{{\cal D}}\thinspace,\nonumber \\
\delta\mathcal{D} & =\mathcal{D}\epsilon_{{\cal J}}^{\prime}+\mathcal{D}^{\prime}\epsilon_{{\cal J}}-\mathcal{P}\epsilon_{{\cal P}}+\mathcal{K}\epsilon_{{\cal K}}+\frac{k}{2\pi}\epsilon_{{\cal D}}^{\prime}\thinspace.\label{eq:Tranflaws}
\end{align}
As pointed out in \cite{Henneaux:2013dra}, the asymptotic form of
the field equations can be obtained from the fact that the evolution
in time corresponds to a gauge transformation spanned by $\Omega=\Omega\left[\mu_{{\cal J}},\mu_{{\cal P}},\mu_{{\cal K}},\mu_{{\cal D}}\right]$,
where $\mu_{{\cal X}}$ stands for the chemical potentials. 

Finally, in order to maintain the fall-off of $a_{t}$, the parameters
$\epsilon_{{\cal X}}$ have to fulfill the following  differential
equations
\begin{align}
\dot{\epsilon}_{{\cal J}} & =\mu_{\mathcal{J}}\epsilon_{{\cal J}}^{\prime}-\epsilon_{\mathcal{J}}\mu_{{\cal J}}^{\prime}+2\left(\epsilon_{\mathcal{P}}\mu_{\mathcal{K}}^{\prime}-\mu_{\mathcal{K}}\epsilon_{\mathcal{P}}^{\prime}-\frac{4\pi}{k}{\cal D}\mu_{\mathcal{K}}\epsilon_{\mathcal{P}}\right)\nonumber \\
 & +2\left(\epsilon_{\mathcal{K}}\mu_{\mathcal{P}}^{\prime}-\mu_{\mathcal{P}}\epsilon_{\mathcal{K}}^{\prime}+\frac{4\pi}{k}{\cal D}\mu_{\mathcal{P}}\epsilon_{\mathcal{K}}\right)\thinspace,\nonumber \\
\dot{\epsilon}_{{\cal P}} & =\mu_{\mathcal{P}}\epsilon_{{\cal J}}^{\prime}-\epsilon_{\mathcal{J}}\mu_{{\cal P}}^{\prime}-\left(\left(\mu_{\mathcal{D}}+\mu_{{\cal J}}^{\prime}\right)\epsilon_{\mathcal{P}}-\mu_{{\cal J}}\epsilon_{\mathcal{P}}^{\prime}\right)+\mu_{{\cal P}}\epsilon_{\mathcal{D}}\thinspace,\label{eq:Epsilon-Punto}\\
\dot{\epsilon}_{{\cal K}} & =\mu_{\mathcal{K}}\epsilon_{{\cal J}}^{\prime}-\epsilon_{\mathcal{J}}\mu_{{\cal K}}^{\prime}+\left(\left(\mu_{\mathcal{D}}-\mu_{{\cal J}}^{\prime}\right)\epsilon_{\mathcal{K}}+\mu_{{\cal J}}\epsilon_{\mathcal{K}}^{\prime}\right)-\mu_{{\cal K}}\epsilon_{\mathcal{D}}\thinspace,\nonumber \\
\dot{\epsilon}_{{\cal D}} & =-\epsilon_{\mathcal{J}}\mu_{{\cal D}}^{\prime}+2\left(\mu_{{\cal K}}^{\prime\prime}-\frac{4\pi}{k}\left({\cal J}\mu_{{\cal K}}+2{\cal D}\mu_{{\cal K}}^{\prime}-\frac{6\pi}{k}{\cal D}^{2}\mu_{{\cal K}}\right)\right)\epsilon_{\mathcal{P}}\nonumber \\
 & -2\left(\mu_{{\cal K}}^{\prime}-\frac{8\pi}{k}{\cal D}\mu_{{\cal K}}\right)\epsilon_{\mathcal{P}}^{\prime}+2\mu_{{\cal K}}\epsilon_{\mathcal{P}}^{\prime\prime}\nonumber \\
 & -2\left(\mu_{{\cal P}}^{\prime\prime}-\frac{4\pi}{k}\left({\cal J}\mu_{{\cal P}}-2{\cal D}\mu_{{\cal P}}^{\prime}-\frac{6\pi}{k}{\cal D}^{2}\mu_{{\cal P}}\right)\right)\epsilon_{\mathcal{K}}\nonumber \\
 & +2\left(\mu_{{\cal P}}^{\prime}+\frac{8\pi}{k}{\cal D}\mu_{{\cal P}}\right)\epsilon_{\mathcal{K}}^{\prime}-2\mu_{{\cal P}}\epsilon_{\mathcal{K}}^{\prime\prime}+\mu_{{\cal J}}\epsilon_{\mathcal{D}}^{\prime}\;.\nonumber 
\end{align}

\end{document}